\documentclass[sigconf, screen]{acmart}
\AtBeginDocument{%
  }

\usepackage{graphicx}
\usepackage{svg}
\usepackage[page]{appendix}
\usepackage{balance}

\copyrightyear{2026}
\acmYear{2026}
\setcopyright{cc}
\setcctype{by}

\begin{document}

\title{AISysRev - LLM-based Tool for Title-abstract Screening}


\author{Aleksi Huotala}
\email{aleksi.huotala@helsinki.fi}
\affiliation{%
  \institution{University of Helsinki}
  \city{Helsinki}
  \country{Finland}}
\orcid{0000-0002-5220-8730}
\author{Miikka Kuutila}
\email{miikka.kuutila@lut.fi}
\affiliation{%
  \institution{LUT University}
  \city{Lahti}
  \country{Finland}
}
\orcid{0000-0002-3695-7280}
\author{Olli-Pekka Turtio}
\email{olli-pekka.turtio@helsinki.fi}
\affiliation{%
  \institution{University of Helsinki}
  \city{Helsinki}
  \country{Finland}
}
\orcid{0009-0004-0820-2958}
\author{Simo Sipilä}
\email{simo.sipila@helsinki.fi}
\affiliation{%
  \institution{University of Helsinki}
  \city{Helsinki}
  \country{Finland}}
\orcid{0009-0003-1398-2613}
\author{Mika Mäntylä}
\email{mika.mantyla@helsinki.fi}
\affiliation{%
  \institution{University of Helsinki}
  \city{Helsinki}
  \country{Finland}}
\orcid{0000-0002-2841-5879}
\renewcommand{\shortauthors}{Huotala et al.}
\acmArticleType{Research}
\acmCodeLink{https://github.com/evotestops/aisysrev}
\acmDataLink{https://zenodo.org/records/17208539}

\begin{abstract}
Conducting systematic reviews is laborious. In the screening or study selection phase, the number of papers can be overwhelming. Recent research has demonstrated that large language models (LLMs) can perform title-abstract screening and support humans in the task. To this end, we developed AISysRev, an LLM-based screening tool implemented as a containerized web application. The tool accepts CSV files containing paper titles and abstracts. Users specify inclusion and exclusion criteria. Multiple different LLMs can be used, such as Gemini, Claude, Mistral or ChatGPT via OpenRouter. We also support locally hosted models and any model compatible with the OpenAI SDK. AISysRev implements both zero-shot and few-shot prompting, and also allows for manual screening through interfaces that display LLM results as guidance for human reviewers. LLM calls are parallelized, meaning screening speed is typically between 100 to 300 papers per minute, depending on the model and the host. To demonstrate the tool's use in practice, we conducted a qualitative trial study with 137 papers using the tool. Our findings indicate that papers can be classified into four categories: Easy Includes, Easy Excludes, Boundary Includes, and Boundary Excludes. The Boundary cases, where LLMs are prone to errors, highlight the need for human intervention. While LLMs do not replace human judgment in systematic reviews, they can reduce the burden of assessing large volumes of scientific literature.
Video: \url{https://www.youtube.com/watch?v=HeblemlgnAQ}
Tool: \url{https://github.com/EvoTestOps/AISysRev}

\end{abstract}




\maketitle

\section{Introduction}
In 2004, Kitchenham et al. \cite{kitchenham2004evidence} published a paper titled "Evidence-Based Software Engineering", which highlights systematic reviews (SRs) as one key element in assembling evidence in software engineering (SE). These ideas originated from medicine~\cite{meade1997selecting} but have since spread to education~\cite{bearman2012systematic}, finance~\cite{kersten2017small} and SE, resulting in a large number of publications adopting the methodology. SRs summarize the state-of-the-art evidence, but conducting them is a laborious task, taking on average 67 weeks to complete~\cite{marshall2019toward}. Title-abstract screening is a phase in SRs where articles are screened against detailed inclusion and exclusion criteria~\cite{kitchenhamGuidelinesPerformingSystematic2007}. The phase may involve thousands of articles -- for example, a number close to 25,000 is reported by Rafi et al.~\cite{rafi2012benefits}.

Given such a large number of studies and the emergence of large language models (LLMs), recent work in SE has started to examine the automation of title-abstract screening~\cite{huotalaPromiseChallengesUsing2024a, felizardoChatGPTApplicationSystematic2024,huotalaSESREvalDatasetEvaluating2025, petersen2025road, thode2025exploring, felizardo2025difficulties}. Our prior work found that LLMs reach up to, and in some cases beyond, the level of master's students, but not up to the level of PhD students ~\cite{huotalaPromiseChallengesUsing2024a}. While initial work demonstrated that we tend to lose some evidence due to false negatives being a problem~\cite{felizardoChatGPTApplicationSystematic2024}, a more recent study reported a recall of 0.98 with a precision of 0.27~\cite{thode2025exploring}. Prompt sensitivity and randomness are highlighted as problems of LLM usage in SE literature reviews~\cite{felizardo2025difficulties}.

An empirical study with 34,528 primary studies from 24 SRs and 9 large language models found that the differences among the top LLMs in SE are relatively small on average, but differences between individual screening tasks are much larger~\cite{huotalaSESREvalDatasetEvaluating2025}. This suggests that the suitability of LLMs for screening should be evaluated on a case-by-case basis. Another motivation for automated tooling is SR variants that emphasize speed, i.e. Rapid reviews that also originated from medicine~\cite{watt2008rapid, garritty2021cochrane}, and later used in SE as well~\cite{pizard2025using}. Even three months was considered to be slow in an industrial setting by a collaborating company~\cite{pizard2025using}, which further emphasizes the need for screening support.

Existing open-source tools (e.g., AIReview~\cite{mao2025aireview}, ASReview~\cite{vandeschootOpenSourceMachine2021} and DenseReviewer~\cite{maoDenseReviewerScreeningPrioritisation2025}) provide AI-assisted support for SRs using LLMs and traditional machine learning methods, but they have practical limitations. In particular, these tools offer limited support for multiple LLM providers~\cite{vandeschootOpenSourceMachine2021, maoDenseReviewerScreeningPrioritisation2025}, have limited documentation~\cite{mao2025aireview} or are no longer actively maintained~\cite{mao2025aireview}. 
Several closed-source commercial tools are also available (e.g., Elicit, Consensus, and Rayyan~\footnote{https://elicit.com/; https://consensus.app/; https://www.rayyan.ai/}), but they limit visibility into the underlying models, parameters, prompts or prompting techniques used.

Taken together, our tool provides research infrastructure that supports large-scale, rapid screening and enables systematic evaluation of screening performance as new LLMs become available. AISysRev is open-source and addresses key challenges, including scalability, parallel execution, heterogeneity among LLM providers, and limitations in documentation and methodological transparency. By offering a graphical user interface (UI), the tool lowers the technical barrier to applying and studying LLM-assisted screening in practice. Thus, the tool provides a foundation for studying and developing guidelines for human-in-the-loop approaches to paper screening in systematic reviews.

\section{AISysRev tool}

\subsection{Design rationale}
The large volume of articles in screening~\cite{rafi2012benefits}, the need for speed and reduced effort~\cite{pizard2025using}, and the substantial variability in LLM-based screening performance across secondary studies~\cite{huotalaSESREvalDatasetEvaluating2025} all motivate our tool AISysRev ~\footnote{https://github.com/EvoTestOps/AISysRev}, a research-based title-abstract screening tool. Our design fully builds on prior findings that LLMs can perform screening at a master’s student level~\cite{huotalaPromiseChallengesUsing2024a}. Our prior work has evaluated over 300,000 screening decisions and compared them against the gold standard \cite{huotalaSESREvalDatasetEvaluating2025}. Thus, the technological foundations of our tool have been established in our prior work.

Both commercial and open-source tools are available for conducting AI-assisted systematic reviews. A major difference between our tool and most commercial ones is that commercial tools typically aim to provide end-to-end support for all SR steps. As a result, there exists the risk where small LLM mistakes may accumulate, with the final output is unlikely to meet scientific quality standards. Our tool’s philosophy follows human-in-the-loop design, with LLMs providing suggestions only.

\begin{figure*}[tbp]
\resizebox{\textwidth}{!}{%
\includegraphics{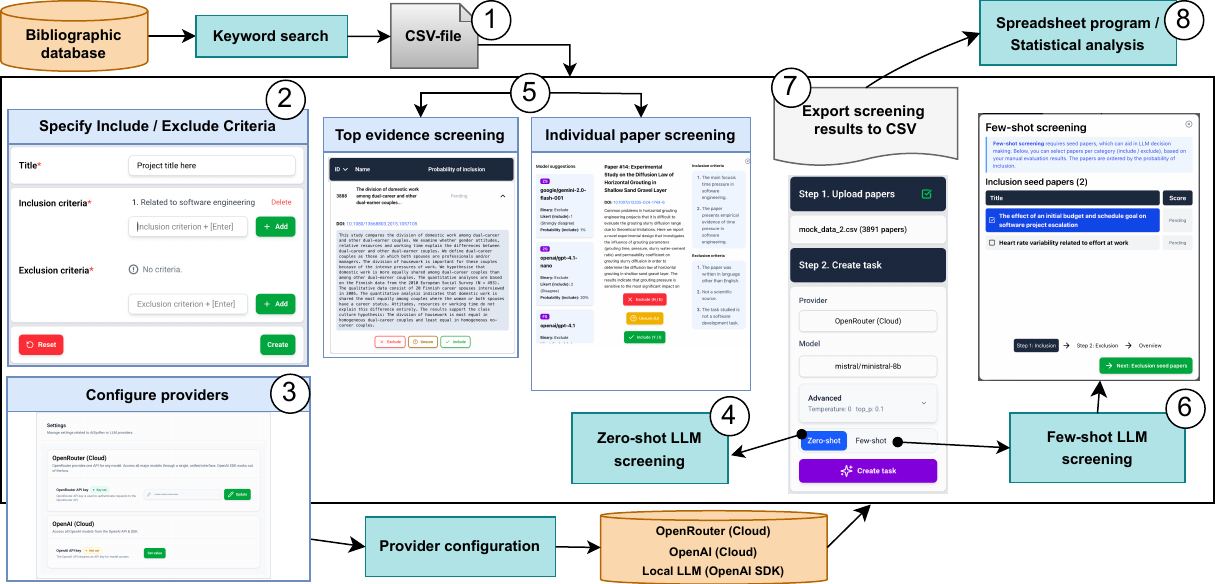}
}%
\Description{Figure showing the AISysRev tool architecture.}
\caption{AISysRev tool architecture.}
\label{figure:aisysrev}
\end{figure*}

Given that LLMs are prone to errors, we designed our tool to feature two primary views (Figure \ref{figure:aisysrev}, part \textcircled{5}). The first view, right side of part \textcircled{5}, allows humans to screen papers individually while also providing recommendations from multiple LLMs. This view is intended for scenarios in which humans plan to screen all papers regardless. In the second view, left side of part \textcircled{5}, papers are ordered based on the probability of inclusion, as computed (averaged) from multiple LLMs. This view is suitable when humans only wish to screen the top evidence, such as in the context of rapid reviews.

\subsection{Feature description}

Main steps of the tool usage are presented in Figure~\ref{figure:aisysrev}.

\subsubsection{Bibliographic database \textcircled{1}}
A bibliographic database (i.e. metadata database), contains paper titles, abstracts, authors, and keywords. Examples of such databases include Scopus, Web of Science, and PubMed. Users provide keywords to search for relevant papers in these databases.

We have implemented and tested AISysRev using Scopus, but since the tool relies on CSV input, it can be easily adapted to work with other databases by modifying the CSV input columns to match that of Scopus. This phase results in a CSV file containing paper titles and abstracts, which serves as the input for the AISysRev tool.

\subsubsection{Setup}
Before papers can be screened, users must specify the inclusion and exclusion criteria \textcircled{2}, which define the rules used by the LLMs to determine whether a paper meets the criteria. Users must also select an LLM provider. AISysRev supports models hosted via OpenRouter, OpenAI, and local OpenAI SDK-compatible services. OpenRouter enables access to LLMs through a unified interface and supports all major providers, including OpenAI, Google, Anthropic, Meta, and Mistral models. Local model support enables on-device LLMs on personal computers or in datacenters. Users can customize LLM behavior on a task-by-task basis, by adjusting parameters such as $temperature$ and $top\_p$ via the UI.

\subsubsection{Screening}
The different prompting techniques for screening can be applied sequentially. The numbering from \textcircled{4} to \textcircled{6} reflects what we consider a typical user workflow. Screening using zero-shot (ZS) prompting \textcircled{4} is the most basic form of LLM-assisted paper screening. In this approach, the LLM evaluates papers based solely on the predefined inclusion and exclusion criteria. To mitigate potential errors, users can perform screening using ZS prompting with multiple models and then compare the results.

Manual screening \textcircled{5} can be conducted in several ways. Users may opt to screen papers manually without any LLM input (left). If LLM-based screening has already been performed, the LLM outputs are displayed in the manual screening window (right). When multiple LLMs are used, the manual screening view shows each LLM’s decision as binary, ordinal, and probability values. The ordinal and probability scores complement the binary decision by highlighting uncertainty, signaling which papers may require human review. Binary, ordinal, and probability values are generated by the LLMs according to the Pydantic schema~\cite{huotalaSESREvalDatasetEvaluating2025}, which defines each classification type.

We have also implemented a list view for manual screening \textcircled{5} (left). This feature is particularly useful for sorting papers based on their probability of inclusion. Users can begin screening with the studies that LLMs identify as most relevant. A cutoff rule can then be applied, e.g., stopping the screening process if ten consecutive papers are excluded. This approach might be valuable in rapid reviews, where speed is prioritized over exhaustive thoroughness.

Screening using Few-shot (FS) prompting \textcircled{6} improves over LLM screening using ZS prompting by incorporating screening decision samples, which have been shown to improve LLM screening performance~\cite{huotalaPromiseChallengesUsing2024a}. This approach involves manually selecting already evaluated papers, both included and excluded, through the UI. These examples are then provided to the LLMs to guide their screening process.

\subsubsection{Output to CSV}
The final LLM and manual screening results can be exported as a CSV file \textcircled{7}. The exported CSV includes basic paper metadata (Title, Abstract and DOI) and columns for a) Every LLM's overall decision; b) Per-criteria decision; and c) Per screening method (ZS or FS) with binary, ordinal and probability values. 

\subsubsection{Spreadsheet program / Statistical analysis \textcircled{8}} The exported CSV can be imported into external spreadsheets or statistical analysis programs. The columns are organized to facilitate the rapid application of spreadsheet formulas, enabling semi-automated decision-making.

\subsection{Technical description}
The AISysRev tool consists of three components: a Python backend, a React-based web frontend, and a Python worker server that handles LLM-specific functionality. Python was selected due to its strong support for asynchronous programming, seamless integration with LLM SDKs via Pydantic AI, and a robust type system that enables a developer-friendly workflow. Pydantic AI automatically corrects errors in structured LLM responses, supports parallelization of LLM calls, and provides strong developer support~\footnote{https://ai.pydantic.dev/}. React.js is a popular JavaScript library with a strong developer and community support. The server implements a REST API, and we provide an OpenAPI specification, if practitioners or researchers wish to use the tool via the API. Nevertheless, we provide a UI focused on usability, accessibility and a rapid screening workflow. AISysRev can run locally or be deployed on cloud platforms such as Amazon AWS, Microsoft Azure, or Google Cloud. The tool runs in a Docker container, using the compose plugin, and its components can be scaled independently based on load.

PostgreSQL is a commonly used relational database system to store and persist data across sessions. In addition, Redis is used as an in-memory cache for real-time and memory I/O intensive actions, such as scheduling OpenRouter API calls using the LLM worker. When screening potentially thousands of papers, I/O performance must be maximized in order to efficiently process them. MinIO is used as the object storage, which shares its API with Amazon's Simple Storage Service (S3) to enable seamless cloud deployment.

The tool is designed to be resilient to common pitfalls in applications that invoke LLMs, such as malformed JSON responses and rate limiting due to high paper screening throughput. If the requests to LLMs fail, we retry them by using an exponential backoff algorithm~\cite{kwak2003analysis}. We parallelize LLM calls with Python's $asyncio$ library~\footnote{https://docs.python.org/3/library/asyncio.html}, which substantially accelerates the screening process. Typically, the tool screens 100 to 300 papers per minute, depending on the model and the host. We found the highest speeds with Google Gemini Flash models.

\section{Tool usage trials}
Our tool has only recently been released, and comprehensive usage studies have not yet been conducted. However, in our past work we have evaluated the prompting techniques (ZS and FS) and internal components (e.g. Python scripts and code) incorporated into the AISysRev tool~\cite{huotalaPromiseChallengesUsing2024a}. The performance of the LLMs in paper screening was demonstrated to be at Master's student level~\cite{huotalaPromiseChallengesUsing2024a}. Furthermore, we benchmarked 34,528 primary studies across nine different LLMs, resulting in a total of over 300,000 screening decision calls via LLM APIs ~\cite{huotalaSESREvalDatasetEvaluating2025}. Token and cost analysis for paper screening was conducted in our previous work~\cite{huotalaSESREvalDatasetEvaluating2025}, which gives a good estimation for token usage and cost.

For this paper, we conducted additional trials with our prior systematic review (SR) on time pressure in software engineering~\cite{kuutila2020time}, which are part of our effort to update our prior SR~\cite{kuutila2020time}. Thus, we have evaluated papers published after the SR was completed. This allowed us to evaluate the tool in an actual context. The trial users (the second and fifth author) were topic experts who had participated in conducting the original SR and can provide gold-standard data on the screening decisions.

\subsection{Trial setup}

In the trials, we used Google Gemini 2.5 Flash, Mistral Small 3.2-23B Instruct and OpenAI GPT-4.1 Mini LLMs. The LLM selection was decided based on a balance of speed, performance and cost. Gemini 2.5 Flash and Mistral Small 3.2-23B Instruct models are smaller and faster compared to the performant but slower GPT-4.1 Mini model. The LLMs were configured to use a temperature parameter of $0.0$, $top\_p$ of $0.1$ and a seed of $128$ to make the results comparable with prior work~\cite{huotalaPromiseChallengesUsing2024a, huotalaSESREvalDatasetEvaluating2025}. Each LLM responded with a binary score, a Likert-scale score (1-7) and a floating point score between 0.000 and 1.000 for both the overall decision and per-criteria basis. For each LLM, we used the ZS prompting technique. The results of the trial run are available as a CSV-file in the research artifact~\cite{zenodo}.

\subsection{Trial results}

These trials yielded qualitative insights regarding the application of our tool for title-abstract screening. A weakness of this approach is that they are based on a single study and may not generalize across SE SRs. In total, we had 137 papers in our screening trials, upon which the following categorization is based. 

\textbf{Easy Includes (n=6)} are those papers in which time pressure is explicitly studied within a software engineering context.
These studies provide empirical evidence and directly address the intersection of time constraints and software development practices, making them highly relevant to our review. We have observed that LLMs include these papers with very high accuracy. For example, \cite{romano2024job} studies the Job Demands and Resources model in the software industry and explicitly describes in the abstract time pressure as a key stressor for software developers.

\textbf{Boundary Includes (n=7)} encompass papers situated within a SE context, but exhibit ambiguity regarding the study of time pressure or the presence of empirical evidence.
While the connection to time pressure may not be explicit, these papers are retained for further examination in the full text screening. It is likely that they often contribute at least partial evidence. For example, a paper on hackathons \cite{sotaquira2025hackathons} includes the following sentences in the abstract: ``hackathons not only enable participants to apply engineering design skills... Additionally, the convergence of competition, time pressure, and an engaging environment creates a stimulating atmosphere that significantly drives skill development and enhances the overall educational experience.'' As the mention of time pressure appears in the abstract after the phrase ``the findings suggest,'' it is implicit that evidence of time pressure in the paper is likely. However, the LLMs often make a mistake and exclude this paper.

\textbf{Boundary Excludes (n=17)} are papers where the task or context is undeniably related to SE, yet the empirical evidence related to time pressure seems very unlikely based on the abstract. Often these papers mention time pressure as motivation or background. For example, \cite{shahzeidi2025hybrid} contains the sentence: "[Technical Debt] problems sometimes arise for reasons such as the time pressure the development team...", which in isolation could indicate that the paper provides evidence on time pressure. However, when examining the context, we found that this sentence is given as background for the main contribution of the paper: "This paper proposes a prediction model called QuantumPSO-LSTM together with the Bidirectional Encoder Representations from Transformers (BERT) word embedding method to identify and classify source code annotations." Having experience of screening thousands of papers on the topic, we believe it is highly unlikely that this paper presents any empirical evidence on time pressure. Nevertheless, LLMs frequently misclassify these papers and provide explanations such as: "The paper discusses time pressure in the context of software engineering, specifically mentioning how time pressure on the development team can lead to technical debt."

\textbf{Clear Excludes (n=107)} consist of studies that, in part, although rigorous in their examination of time pressure, do not pertain to SE tasks. For instance, research conducted in fields such as medicine or aviation fall outside the scope of our review. In our SR on time pressure, there exists a difficulty in effectively filtering relevant papers using keyword-based searches, resulting in a substantial number of papers being easy excludes. Both LLMs and humans accurately exclude these papers, and it is indeed the volume of these papers where LLMs could make a useful contribution.

\balance

\section{Conclusions and Future work}

This paper introduced AISysRev, a state-of-the-art LLM-based tool for title-abstract screening that bridges the gap between research and practical tooling. The tool supports both ZS and FS prompting techniques with local or cloud-hosted LLMs, while allowing control over model selection and parameters. AISysRev also supports exporting screening results in CSV format for further statistical analysis in other tools. To the best of our knowledge, no existing LLM-based screening tool offers comparable flexibility or transparency in terms of prompting techniques, multi-provider LLM access, scalability, and support for locally hosted models. In contrast, commercial tools (e.g. Elicit, Consensus, and Rayyan) are typically closed source, which limits the visibility into models, parameters, and prompting techniques. Our tool is open-source, which promotes transparency and continued development of the tool.

The underlying screening technology of AISysRev was evaluated in our prior work~\cite{huotalaSESREvalDatasetEvaluating2025}, and the tool itself was tested using trial runs, where we found that papers can be classified into Easy Includes, Easy Excludes, Boundary Includes, and Boundary Excludes. The boundary cases highlight the need for human intervention. While we evaluated the tool purely within papers in the SE domain, the AISysRev tool itself is non-domain-specific. We encourage practitioners to test the tool also in domains other than SE.

In the future, we plan to compare the tool's performance to other AI-powered assistants, such as Elicit, Consensus or Rayyan. We also plan to expand the feature set of AISysRev - including expanding the tool to search strategy generation and keyword identification~\cite{kitchenhamGuidelinesPerformingSystematic2007}, database search, full-text screening, data extraction, and using embeddings to visualize the decision boundary. The usefulness of the tool could also be demonstrated with user studies comparing the screening performance of human raters with and without the tool.

\begin{acks}
The first, second and fifth author have been funded by the Strategic Research Council of Research Council of Finland (Grant ID 358471). 
\end{acks}

\bibliographystyle{ACM-Reference-Format}
\bibliography{acmart}

\end{document}